\documentclass[epj,online]{webofc}
\fancypagestyle{firstpage}{%
  \fancyhf{}%
  \fancyhead[R]{%
    \minipage[t][\baselineskip]{\linewidth}
      \small
      \emph{Proc. CONF12} -- XII$^{\rm th}$ Quark Confinement and the Hadron Spectrum \hfill CERN-TH-2016-237\\
      to appear in EPJ Web of Conferences \hfill IFT-UAM/CSIC-16-120\\
    \endminipage
  }%
}
\usepackage[utf8]{inputenc}
\usepackage[varg]{txfonts}   
\usepackage{xcolor}
\definecolor{darkred}{rgb}{0.4,0.0,0.0}
\definecolor{darkgreen}{rgb}{0.0,0.4,0.0}
\definecolor{darkblue}{rgb}{0.0,0.0,0.4}
\usepackage[bookmarks,linktocpage,colorlinks,
    linkcolor = darkred,
    urlcolor  = darkblue,
    citecolor = darkgreen]{hyperref}
\wocname{EPJ Web of Conferences}
\woctitle{CONF12}
\newcommand{\Nf}{\ensuremath{N_{\rm f}}}
\newcommand{\MSbar}{\ensuremath{\kern1pt\overline{\kern0pt \rm MS\kern-1pt}\kern1pt}\xspace}
\newcommand{\flav}{flavour\xspace}
\newcommand{\gbar}{\kern1pt\overline{\kern-1pt g\kern-1pt}\kern1pt}
\newcommand{\gbsq}{\gbar^2}
\newcommand{\gbGF}{\gbar_{\rm GF}}
\newcommand{\gbSF}{\gbar_{\rm SF}}
\newcommand{\uSF}{u_{\rm SF}}
\newcommand{\ZP}{{Z_{\rm P}}}
\newcommand{\ZA}{{Z_{\rm A}}}
\newcommand{\ZM}{{Z_{\rm M}}}
\newcommand{\fm}{{\rm fm}}
\newcommand{\ee}{{\rm e}}
\newcommand{\dd}{{\rm d}}
\newcommand{\GeV}{{\rm GeV}}
\newcommand{\MeV}{{\rm MeV}}
\newcommand{\Lhad}{L_{\rm had}}
\newcommand{\Lmax}{L_{\rm max}}
\newcommand{\Lmin}{L_{\rm min}}
\newcommand{\lswi}{L_{\rm swi}}

\newcommand{\muhad}{\mu_{\rm had}}
\newcommand{\mupt}{\mu_{\rm PT}}
\newcommand{\muswi}{\mu_{\rm swi}}
\newcommand{\gsim}{\mathrel{\mathop\sim^{g\to 0}}}
\newcommand{\mbar}{\kern1pt\overline{\kern-1pt m\kern-1pt}\kern1pt}
\newcommand{\sigP}{\sigma_{\rm P}}
\newcommand{\mpiz}{\ensuremath{m_{\tiny\pi^0}}}
\newcommand{\mKz}{\ensuremath{m_{\tiny\rm K^0}}}
\newcommand{\mDz}{\ensuremath{m_{\tiny\rm D^0}}}
\newcommand{\mBz}{\ensuremath{m_{\tiny\rm B^0}}}
\newcommand{\mZ}{\ensuremath{m_{\tiny\rm Z}}}

\newcommand{\vp}[1][\vphantom{'}]{^{#1}}
\newcommand{\alphas}{\alpha_{\rm s}}
\newcommand{\psibar}{\overline{\psi}}
\newcommand{\tr}{\,\hbox{tr}\,}
\begin{document}
%
\selectlanguage{english}
\title{%
Controlling quark mass determinations non-perturbatively in three-flavour QCD 
}
\author{
Isabel Campos\inst{1} \and 
Patrick Fritzsch\inst{2,4}\fnsep\thanks{\email{p.fritzsch@csic.es}} \and
Carlos Pena\inst{2,3} \and
David Preti\inst{2} \and
Alberto Ramos\inst{4} \and
Anastassios Vladikas\inst{5}
}
\institute{%
Instituto de F\'{\i}sica de Cantabria / Instituto de F\'{\i}sica Te{\'o}rica (IFCA/IFT-CSIC)
\and
Instituto de F\'{\i}sica Te{\'o}rica UAM/CSIC, Universidad Aut{\'o}noma de Madrid, \\
C/ Nicol{\'a}s Cabrera 13-15, Cantoblanco, Madrid 28049, Spain
\and
Universidad Aut{\'o}noma de Madrid, Cantoblanco, Madrid 28049, Spain
\and
Theoretical Physics Department, CERN, 1211 Geneva 23, Switzerland
\and
INFN, Sezione di Tor Vergata, c/o Dipartimento di Fisica, Università di Roma Tor Vergata, \\
Via della Ricerca Scientifica 1, 00133 Rome, Italy
}
\abstract{%
The determination of quark masses from lattice QCD simulations requires a
non-perturbative renormalization procedure and subsequent scale evolution to
high energies, where a conversion to the commonly used \MSbar
scheme can be safely established. We present our results for the
non-perturbative running of renormalized quark masses in $\Nf=3$ QCD
between the electroweak and a hadronic energy scale, where lattice simulations
are at our disposal. Recent theoretical advances in combination with
well-established techniques allows to follow the scale evolution to very high
statistical accuracy, and full control of systematic effects. 
}
%
\maketitle
%
\section{Quantum Chromodynamics at all scales}
\label{intro}

\begin{figure}[th]
  \small
  \centering
  \includegraphics[width=\textwidth]{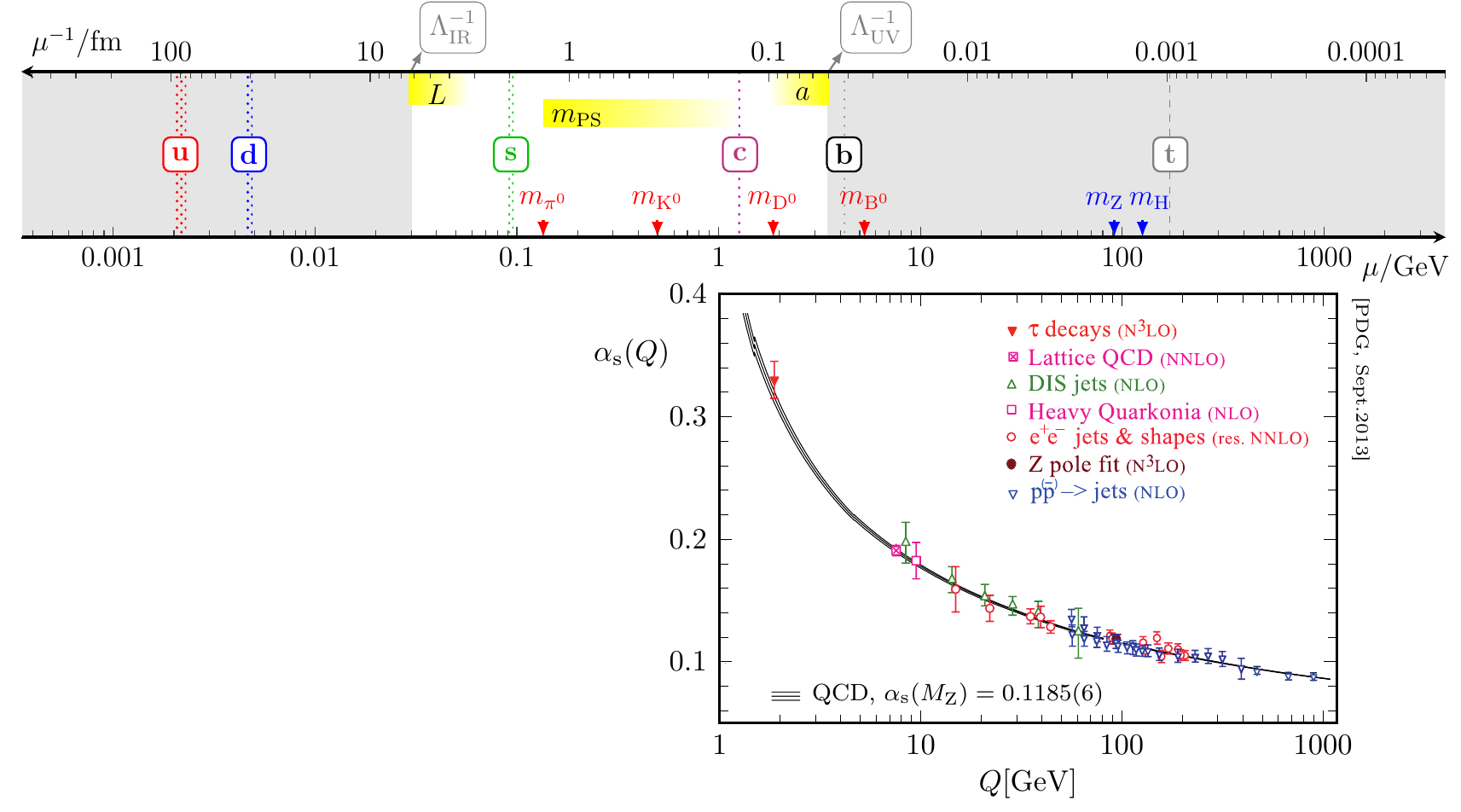}
  \caption{Sketch of energy scales relevant to QCD phenomenology. The masses
           for the six quark \flav{s} found in nature so far
           (u,\,d,\,s,\,c,\,b,\,t) are taken as quoted in~\cite{PDG:2014}. Also
           the masses of the neutral pseudoscalar meson bound states
           ($\mpiz$,\,$\mKz$,\,$\mDz$,\,$\mBz$) as well as the depicted scale
           dependence of the strong coupling $\alphas$ are taken from that
           reference. Although, $\alphas(\mu)$ can be computed in any sensible
           renormalization scheme one conveniently quotes it for the 5-\flav
           theory at the electroweak scale $\mu=\mZ$ in the \MSbar
           scheme.\newline 
           The lattice regularisation of QCD provides the only known
           (non-perturbative and gauge-invariant) framework to compute bound
           states of the strong interaction. For phenomenological applications
           one is restricted to a window ($\Lambda_{\rm
           IR}\le\mu\le\Lambda_{\rm UV}$) in the low-energy regime of QCD in
           order to incorporate the important long-distance effects. They are
           of the order of the Compton wavelength of the lightest particle,
           $\lambda_{\pi^{0}}=1/\mpiz$. It has to be well separated from the
           infrared cutoff $\Lambda_{\rm IR}^{-1}\sim L \gg \lambda_{\pi^{0}}$
           in order to not be distorted by finite volume effects. Typical
           simulations have a physical extent of $L\gtrsim 3\,\fm$ and a
           number of lattice points of $N=(L/a)^4 \approx (50-100)^4$, thus
           restricting the lattice spacing to $a\sim\Lambda_{\rm
           UV}^{-1}\gtrsim0.045\,\fm$. To extract physical quantities one also
           has to stay away from the ultraviolet cutoff, i.e.
           $\mu\ll\Lambda_{\rm UV}$, otherwise control over finite lattice
           spacing effects is lost, and the continuum limit $a\to 0$ cannot
           be taken. Although, various advances in LQCD steadily increase the
           quality and reliability of numerical simulations with different
           number of \flav{s} $\Nf$, the cost of simulations and algorithmic
           difficulties still prevent us from reaching even smaller lattice
           spacings. \newline
           For any non-perturbative renormalization problem posed at fixed
           renormalization scale $\mu$ in the low-energy regime of LQCD the
           continuum limit has to be taken in a controlled way. Then the
           remaining question is how to relate the obtained results to
           experiments at much higher energies, typically two orders of
           magnitude. For high precision physics the supposedly most convenient
           renormalization scheme, \MSbar, is of little help at scales much
           below $10\,\GeV$ (or $\alphas\gtrsim 0.2$). Its intrinsic
           perturbative nature, the behaviour of asymptotic series expansions
           in general, and missing non-perturbative contributions prevent us
           from quantifying its (non-)reliability at low energies from within
           the scheme itself. As detailed in the main text the renormalization
           group running of any operator can be determined purely
           non-perturbative, thus connecting the low-energy regime of LQCD in
           a controlled and quantifiable way to scales well above $10\,\GeV$.
           Recently, it has become possible to non-perturbatively test the
           accuracy of continuum perturbation theory for $\alphas\lesssim
           0.2$~\cite{Brida:2016flw}, stressing the relevance and necessity of
           a careful assessment beyond perturbation theory in present high
           precision physics. Non-perturbatively it is even possible nowadays
           to follow the renormalization group in QCD down to scales of about
           $200\,\MeV$, cf. Ref.~\cite{DallaBrida:2016kgh}.
           Both results have been reported at this
           conference~\cite{Sint:12Conf2016,DallaBrida:12Conf2016}.
          }
  \label{fig:RGscale}
\end{figure}

For numerical simulations one starts from one of many valid discretisations
of the (bare) Euclidean action density of Quantum Chromodynamics (QCD), 
\begin{align}\label{eq:QCDact}
  \mathcal{L}_{\tiny QCD} &=  
       \frac{1}{2g_0^2}\tr\{ F_{\mu\nu} F_{\mu\nu} \} 
     + \sum_{i=1}^{\Nf} \psibar_i \big( \gamma_\mu D_\mu + m_{0,i} \big) \psi_i  \;, & 
     D_{\mu} &= \partial_{\mu} + A_{\mu}  \;,
\end{align}
with a fixed number of dynamical quark \flav{s} $\Nf$. With $\Nf+1$ input
parameters the action is supposed to describe the strong interaction at all
energy scales. Lattice QCD is a gauge-invariant regularisation and thus does
not require a gauge-fixing and Faddeev--Popov term. It is the only
regularisation allowing to determine colorless bound states
$(\mpiz,\mKz,m_{\rm p},\ldots)$ between quarks from first principle
calculations, which due to computational and theoretical advances can be
systematically improved over time. However, incorporating all physical
mass-scales that are relevant for describing ($\Nf=6$) QCD---as it seems to be
realised in nature---is an impossible task to present state-of-the-art
simulations due to the immense costs to accommodate many orders of magnitude of
energy scales in a single simulation, cf. Figure~\ref{fig:RGscale}. Being
mainly interested in non-perturbative, long-distance effects of QCD, the 
phenomenological applicability of LQCD is limited to a fixed window%
\footnote{%
for convenience, we shall refer to all physical scales well separable from the
cutoffs as 'low-energy' window in the following, i.e., $\mu \ll 1/a \sim 2-4\,\GeV$
}
\begin{align}\label{eq:LEW}
        40\,\MeV                                      \;\sim\; 
        \Lambda_{\tiny\rm IR}                         \;\ll\; 
        \muhad,\,\mpiz,\, m_{\rm p},\, \mKz,\, \ldots \;\ll\;
        \Lambda_{\tiny\rm UV}                         \;\sim\; 
        3\,\GeV  \;.
\end{align}
Any physically interesting scale $\mu$ has to be well separated from the
imposed infrared and ultraviolet cutoffs, $\Lambda_{\tiny\rm IR} \ll \mu
\ll\Lambda_{\tiny\rm UV}$, to allow for a controlled assessment of the
respective cutoff effects, see caption of Figure~\ref{fig:RGscale}.  Although,
the lattice community can afford to simulate with three or four dynamical
\flav{s} nowadays, one still needs to balance the $\Nf+1$ chosen (bare) input
parameters, given in terms of $\Nf+1$ physical scales from the real spectrum,
against the intrinsic cutoff values and algorithmic costs.  As no practical
perfect lattice setup exists, different views on how to best achieve this
balance lead to controversies among lattice practitioners.  Another of those
disputable topics is the application of perturbation theory at energies as low
as those given by the low-energy window of lattice QCD as specified
in~\eqref{eq:LEW}, including $\mu=\Lambda_{\tiny\rm UV}\sim\,3\,\GeV$, see
also~\cite{Brida:2016flw}.

In course of our determination of quark masses, we have to fix a
renormalization prescription at scale $\muhad$ that lies well inside the window
$[\Lambda_{\tiny\rm IR},\Lambda_{\tiny\rm UV}]$.  Then the natural question
arises how to compare or connect our result to other determinations as
summarised for instance by the Particle Data Group~\cite{PDG:2014}. Both
naturally differ in the renormalization scheme and scale.  After removing the
UV cutoff dependence, we thus have to evolve our (continuum) result,
$\mbar_{i}(\muhad)$, using renormalization group (RG) transformations to the
same physical scale and scheme, typically \MSbar.

In a mass-independent renormalization scheme for QCD, the renormalization group
equations (RGE) for the \textit{running coupling} and \textit{running quark
mass(es)} read
\begin{align}\label{eq:RGeq}
                    \mu\dfrac{\partial}{\partial \mu}\gbar(\mu) &=\beta(\gbar) \;, &
     \dfrac{\mu}{\mbar}\dfrac{\partial}{\partial \mu}\mbar(\mu) &=\tau (\gbar) \;,
\end{align}
and can be formally integrated. The resulting, exact solutions are known as
renormalization group invariants (RGI), conveniently written as
\begin{align}  \label{eq:Lambda}
   {\Lambda} &\equiv  
                         \mu\big[{b_0\gbsq(\mu)}\big]^{-{b_1}/({2b_0^2})}\,\ee^{-1/({2b_0\gbsq(\mu)})} 
                         \exp\bigg(\!-\!\!\int_{0}^{\gbar^{\vphantom{A}}(\mu)}\!\!\!\!\dd{g}\,
                         \bigg[\dfrac{1}{\beta(g)}+\dfrac{1}{b_0g^3} -\dfrac{b_1}{b_0^2g}\bigg]\bigg)  \;, \\[0.5em] 
               \label{eq:Mi}
      { M_i} &\equiv  
                         \mbar_i(\mu)\big[{2b_0\gbsq(\mu)}\big]^{-{d_0}/({2b_0})}\,
                         \exp\bigg(\!-\!\!\int_{0}^{\gbar^{\vphantom{A}}(\mu)}\!\!\!\!\dd{g}\,
                         \bigg[\dfrac{\tau(g)}{\beta(g)}-\dfrac{d_0}{b_0g}\bigg]\bigg)  \;, \quad i\in\{\rm u,d,s,c,b,t\} \,.
\end{align}
Only the leading, universal (scheme-independent) coefficients $b_0, b_1, d_0$
of the perturbative expansions of the beta-function and mass-anomalous
dimension,
\begin{align}\label{eq:anomalous-PT}
    \beta(g)  &\gsim  -g^3({ b_0}+{ b_1}g^2+{ b_2^s}g^4+\ldots) \;,  & 
    \tau (g)  &\gsim  -g^2({ d_0}+{ d_1^s}g^2+\ldots)           \;, 
\end{align}
appear such that the individual integrands are finite when $g\to 0$.
Higher order coefficients are scheme-dependent---indicated by the superscript
$s$---and so are $\beta$ and $\tau$. Thanks to recently completed efforts these
approximations are now known to 5-loop order in the minimal subtraction
scheme(s), cf.~\cite{Gross:1973id,Politzer:1973fx,Caswell:1974gg,Tarasov:1980au,vanRitbergen:1997va,Czakon:2004bu,Baikov:2016tgj}
and~\cite{Tarrach:1980up,Tarasov:1982gk,Chetyrkin:1997dh,Vermaseren:1997fq,Baikov:2014qja}.
However, it is well-known that by using an asymptotic series expansions one is
missing non-perturbative contributions%
\footnote{%
typically associated with instantons, renormalons, \ldots
}
which become increasingly more relevant towards low energies, i.e., when the
strong coupling becomes large.  One has to appreciate that these issues can
be overcome by determining $\beta(g)$ and $\tau(g)$ non-perturbatively.
In that case, eqs.~\eqref{eq:Lambda} and \eqref{eq:Mi} determine the
fundamental parameters $\{\Lambda,M_i\}_{\Nf}$ uniquely for any input scale
$\mu$.

\clearpage

\section{An intermediate non-perturbative renormalization scheme}

\begin{figure}[t]
  \small
  \centering
  \includegraphics[width=0.7\textwidth]{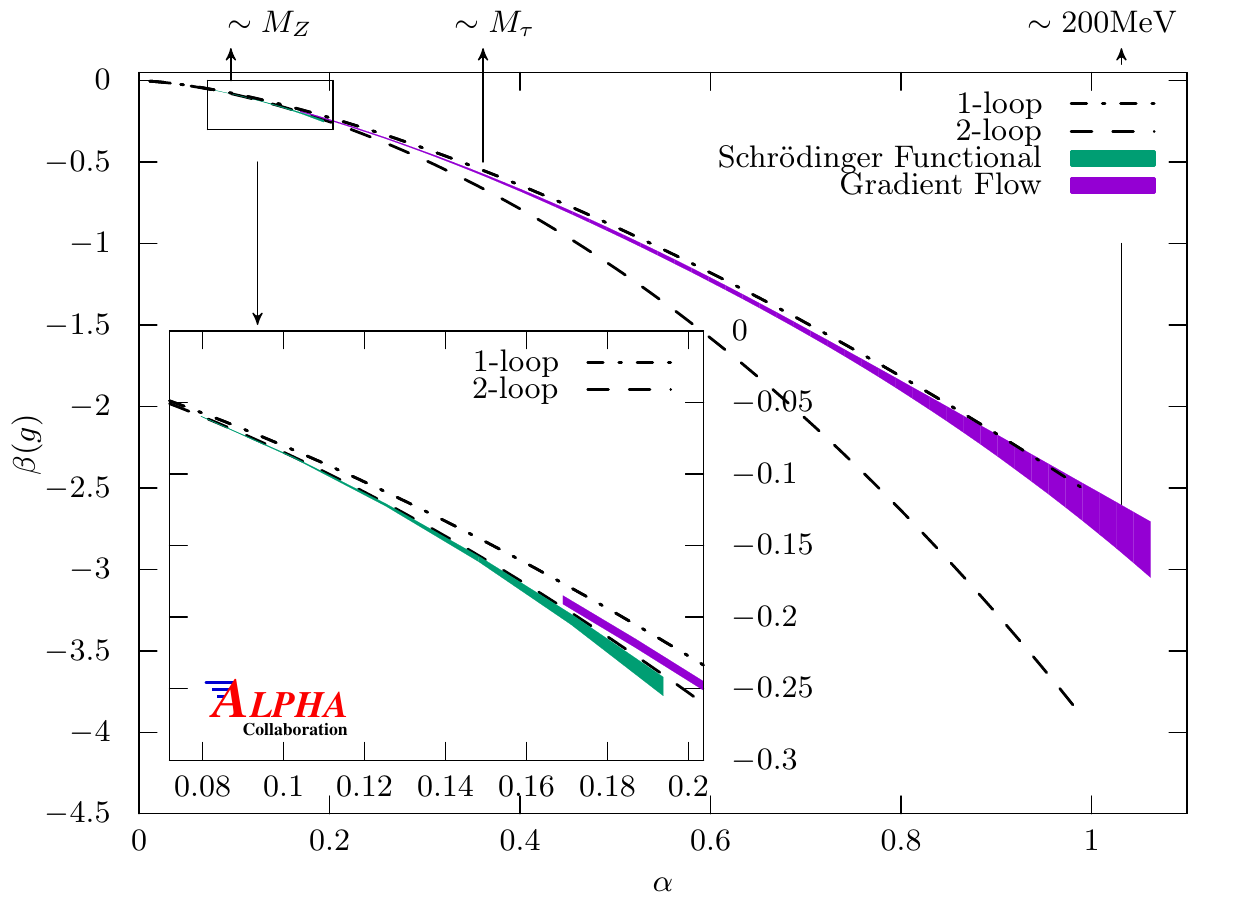}
  \caption{Non-perturbative $\beta$-function for the SF 
           ($\alpha_{\rm SF}\lesssim 0.2$) and GF 
           ($\alpha_{\rm GF}\gtrsim 0.17$) running coupling scheme; picture
           from~\cite{DallaBrida:2016kgh}. There also the two schemes have been
           matched non-perturbatively at fixed physical volume corresponding to
           about $4\,\GeV$.
          }
          \label{fig:betas}
\end{figure}

For massless renormalization schemes (RS) it is most natural to first solve the
RG equation for the coupling and then for the mass, cf.~\eqref{eq:RGeq}. In
LQCD it has become customary to determine the scale evolution of coupling and
mass using a \emph{finite-volume renormalization scheme} in combination with
recursive finite-size scaling. For that purpose the physical size $L$ of the
simulated volume $L^4$ is identified with the inverse renormalization scale for
the sole purpose of solving the RG equations by rescaling the box size, $L\to sL$.
The constant factor $s$ is usually chosen to be $2$. In that way one is able
to cleanly disentangle the renormalization procedure from the low-energy window
of LQCD and to connect the hadronic renormalization scale $\muhad=1/\Lhad$ to
the electroweak scale $\mu=\mZ$, where for instance a change of renormalization
schemes towards the commonly quoted \MSbar scheme can typically be achieved in
a controlled way,%
\footnote{%
for the time being we stay at fixed $\Nf$ and neglect changes from $\Nf=3\to 5$
}
cf. Figure~\ref{fig:RGscale}.
For that purpose one should ideally cover a range of scales compatible with
\begin{align}\label{eq:running}
        \Lmax &\ge \Lhad \;, &   \Lmin &\le 1/\mZ \;, & \Lmax/\Lmin &= s^{N_s} \;, & s>1 \;.
\end{align}
For example, to connect a scale $\muhad$ as low as $200\,\MeV$ to the
electroweak scale of about $100\,\GeV$ using $s=2$, at least $N_s=9$ individual
steps are necessary.  Each of the $N_s$ steps requires a series of lattice
simulations at different resolutions $L/a$ but fixed renormalized input
parameters, $\gbsq(L)=u$ and $L\mbar_i(L)=0$. In this way $\mu=1/L$ is kept
fixed while only the lattice spacing $a$ is being varied, or equivalently, the
bare parameters $\{g_0^2,m_{0,i}\}$ as a function of $a$. A second set of
simulations at the same bare parameters but $s L/a$ number of points in each
direction permits to trace the change of renormalization scale $L\to sL$ at
finite $a$. With adequate renormalization conditions for the coupling and quark
mass, this allows to take the continuum limit $a\to 0$ of their lattice
approximants in order to determine the response w.r.t. a change of scale $s$ in
the continuum, equivalent to
\begin{align}\label{eq:ssf}
        -\ln(s) &=  \int^{\,\gbar(sL)}_{\,\gbar(L)} \!\!            \frac{{\rm d}g}{\beta(g)}  \;, &
     \ln(\sigP) &= -\int^{\,\gbar(sL)}_{\,\gbar(L)} \!\!\! {\rm d}g \frac{ \tau(g)}{\beta(g)}  \;, &
          \text{for fixed} &\quad \begin{cases} \;\;\;\gbsq(L) =u \\ \,L\mbar_i(L) =0  \end{cases} \;.
\end{align}
In this way, the relevant information is encoded in the \emph{step-scaling functions} 
(SSFs)~\cite{Luscher:1993gh,Capitani:1998mq}, 
\begin{align}\label{eq:SSF}
   \sigma(s,u) &= \gbsq(sL)  \;, & 
   \sigP (s,u) &= \mbar_{i}(L) \big/ \mbar_i(sL) \;.
\end{align}
While various ways exist to determine the SSFs numerically via lattice
simulations of QCD, the so-called \emph{Schr\"odinger functional} (SF)
setup~\cite{Luscher:1992an,Sint:1993un,Sint:1995rb} has been develop for
exactly that purpose. For instance, in contrast to standard (anti-)periodic
boundary conditions in all space time directions, Dirichlet boundary conditions
are imposed in time direction. This allows for a natural non-perturbative
definition of a strong coupling via non-vanishing boundary gauge fields as well
as to simulate at vanishing quark mass. For additional details we point to the
review article~\cite{Sommer:2015kza} and references therein.

The scale evolution of the strong coupling has been determined in
refs.~\cite{Brida:2016flw,DallaBrida:2016kgh}, and a short account of the whole
procedure can be found in~\cite{Bruno:2016gvs}.  However, a technical
complication is inherited from that determination: instead of a single RS, two
different schemes are being used and matched non-perturbatively at a scale of
approx.  $4\,\GeV$. At present, its physical motivation is two-fold: 1) control
the perturbative matching at the electroweak scale to high
precision~\cite{Brida:2016flw} using the Schr\"odinger functional coupling,
$\gbSF$, and 2) reach good statistical accuracy down to energies of about
$200\,\MeV$~\cite{DallaBrida:2016kgh} by applying the Gradient flow coupling
scheme, $\gbGF$~\cite{Fritzsch:2013je,Ramos:2015baa}.  Compared to previous
estimates of $\alphas(\mZ)$ this represents a new quality of rigor from lattice
QCD determinations and for the first time allowed to accurately determine the
corresponding non-perturbative QCD $\beta$-functions, see
Figure~\ref{fig:betas}. 

For our determination of the scale evolution of renormalized quark masses in
three-flavour QCD, this does not pose any additional obstacles but only
complicates the overall presentation. In fact, assuming a diagonal quark mass
matrix of rank $\Nf$ in two massless schemes, one can write~\cite{Sint:1998iq}
\begin{align}
  \mu'        &= c\mu\,,   \; c > 0, &
  \gbar'^2    &= \gbsq           {\cal X}_{\rm g}(\gbar)\,, &
  \mbar_{j}'  &= \mbar_{j}\vp \, {\cal X}_{\rm m}\vp(\gbar)\,, & j=1,\ldots,\Nf \,.
\end{align}
Invariance of a physical observable $P$ under this change of variables implies 
\begin{align}
       P' \big[\,\mu'(\mu),\,\gbar'(\gbar),\,\{\mbar_{j}'(\gbar,\mbar_{j}\vp)\}\,\big]
    &= P\,\big[\,\mu      ,\,\gbar        ,\,\{\mbar_{j}\}                    \big]  \;,
\end{align}
where $P'$ satisfies the Callan--Symanzik equation in the primed scheme w.r.t.
\begin{align}
  \beta'(\gbar')&= \left\{\beta(\gbar)\frac{\partial\gbar'}{\partial\gbar} \right\}_{\gbar=\gbar(\gbar')}
   \;, & 
  \tau'(\gbar') &= \left\{\tau(\gbar)+\beta(\gbar)
                     \frac{\partial}{\partial\gbar}\ln{\cal X}_{\rm m}(\gbar)
                   \right\}_{\gbar=\gbar(\gbar')}  \;.
\end{align}
While this is the general connection between two massless schemes at the level
of RG functions, every step in our determination is done non-perturbatively.
In practise, when switching between the two schemes, we do not change the
imposed renormalization condition---which is relevant for determining
$\sigP(2,u)$---but merely the implicit, parametric dependence on the
renormalized coupling $u\equiv\gbsq(L)$. The connection between SF and GF
schemes has been established non-perturbatively at a well-chosen fixed physical
box size, $L_0\approx (4\,\GeV)^{-1}$, and reads~\cite{DallaBrida:2016kgh}
\begin{align}\label{eq:u_match}
   \gbSF^2(L_0)  &= 2.012      \;, &
   \gbGF^2(2L_0) &= 2.6723(64) \;. 
\end{align}

\section{The running quark mass}

\begin{figure}[t]
  \small
  \centering
  \includegraphics{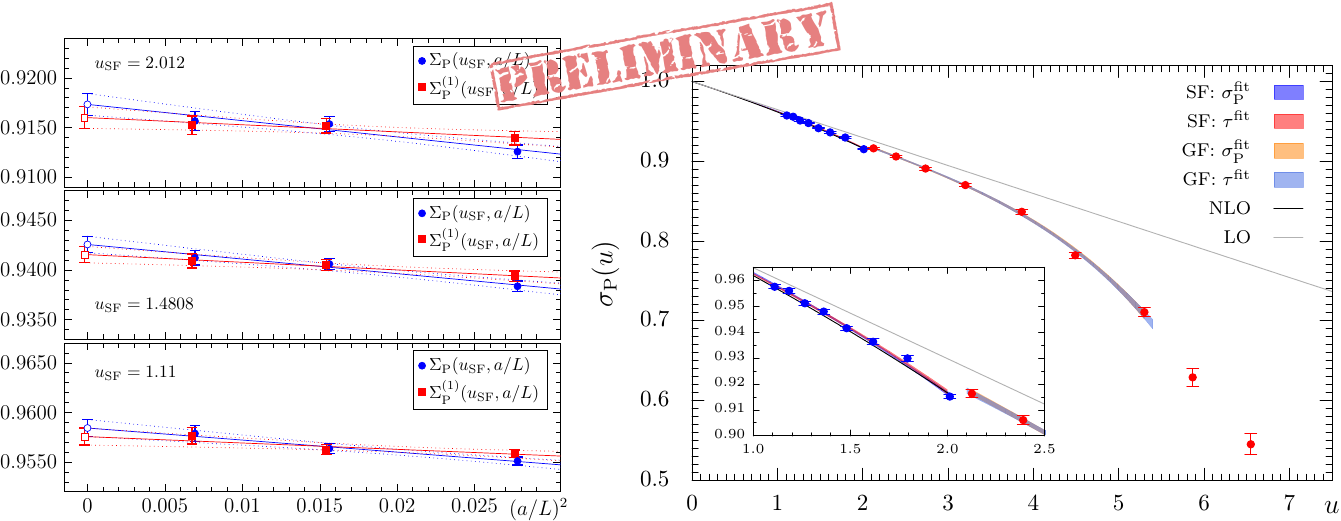}
  \caption{\emph{Left:} Representative continuum extrapolations ($u$-by-$u$) for
           the step-scaling function $\sigP(u)\equiv\lim_{a/L\to 0}\Sigma^{(1)}(u,a/L)$ 
           in the SF scheme. For further details please consult~\cite{Campos:2015fka,Lattice16}.
           \emph{Right:}
           Non-perturbatively determined step-scaling functions $\sigP(u)$ in
           the SF ($u\lesssim 2.0$) and GF ($u\gtrsim 2.1$) running coupling 
           scheme, together with different interpolating fits. We note that for 
           $\uSF\lesssim 1.5$ the data agrees with NLO perturbation theory 
           such that the matching with SF-PT can be safely established.
          }
          \label{fig:ssf}
\end{figure}

Given the discussions in the previous sections, our strategy of a
non-perturbatively controlled determination of quark masses in the
three-flavour theory with two renormalization schemes proceeds as follows. We
determine the flavour-independent RG factor ${M}/{\mbar(\muhad)}$, cf.
eq.~\eqref{eq:Mi}, that connects the RGI mass to the quark mass in the
hadronic, low-energy regime 
\begin{align}\label{eq:masterF}
   \frac{M}{\mbar(\muhad)} &= \left[\frac{\mbar(\muswi)}{\mbar(\muhad)}\right]_{\rm GF-NP} \!\!\times\;
                              \left[\frac{\mbar(\mupt) }{\mbar(\muswi)}\right]_{\rm SF-NP} \!\!\times\;
                              \left[\frac{M}{\mbar(\mupt)}             \right]_{\rm SF-PT} \;.
\end{align}
Note that we intend to make use of perturbation theory at a scale $\mupt$ close
to $\mZ$ where truncation errors from using the known 3-/2-loop orders in the
perturbative $\beta$-/$\tau$-functions in the SF scheme are sufficiently 
suppressed,%
\footnote{%
  at high energies the (non-perturbative) SF scheme is parametrically close to
  the (perturbative) $\MSbar$ scheme
}
see Figure~\ref{fig:ssf}. The true challenge is to determine the two
non-perturbative factors to a sufficient precision while controlling systematic
effects at the sub-percent level for each individual contribution. 
When this is achieved, the scale- and scheme-independent RGI mass of 
flavour $j$ can be determined from
\begin{align}\label{eq:hadMatch}
        M_j  &= \frac{M}{\mbar(\muhad)} \times \mbar_{j}(\muhad) \;,
\end{align}
where $\mbar_{j}(\muhad)$ constitutes a renormalized quark mass
computed in the hadronic regime. Ideally, $M_j$ is the fundamental 
parameter to be compared to other determinations. However, it has
become customary to quote \MSbar masses. In order to do so one 
subsequently has to employ perturbation theory and, depending on
the individual flavour and renormalization scale, appropriately 
match at the charm and bottom thresholds.
\begin{figure}[t]
  \small
  \centering
  \includegraphics[width=0.7\textwidth]{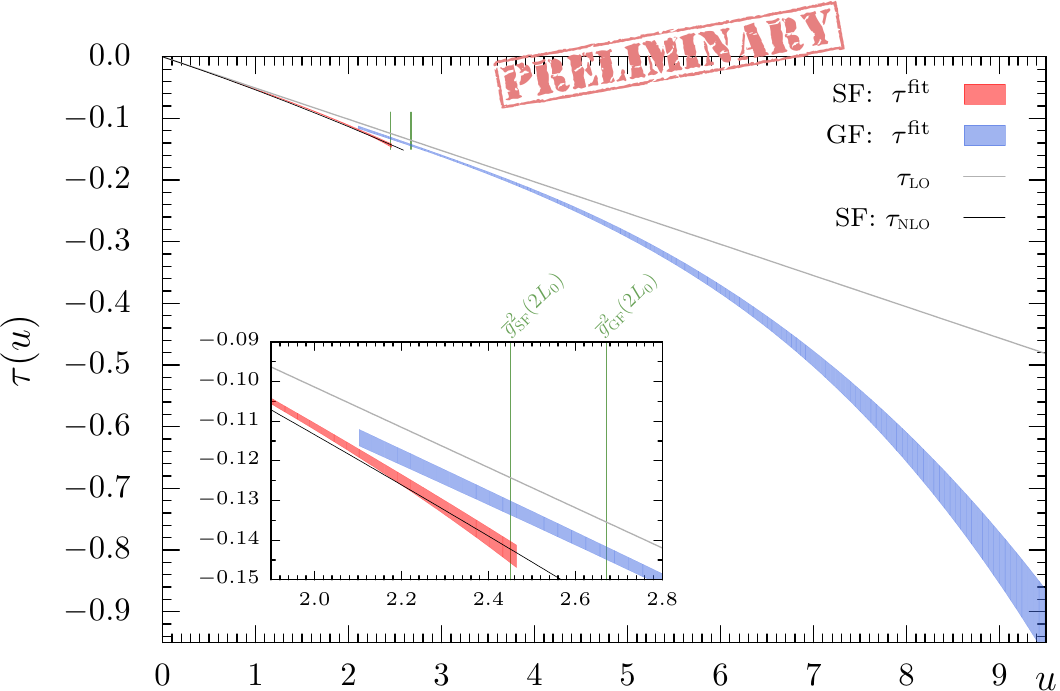}
  \caption{Non-perturbative mass anomalous dimension $\tau(g^2)$ in the SF 
           ($\gbsq_{\rm SF}\lesssim 2.45$) and GF ($\gbsq_{\rm GF}\gtrsim 2.10$) 
           running coupling scheme. The vertical lines in the inner plot 
           represent the respective couplings at the switching scale. Note
           that for $u=\gbGF^2$ our data exactly covers the plotted fit range.
          }
          \label{fig:tau}
\end{figure}
An important fact worth emphasising is that the continuum running factor in
eq.~\eqref{eq:masterF} can be used by other lattice practitioners which may
employ a different lattice discretization in their computations. They just have
to recompute $\mbar_{j}(\muhad)$ using the same---but so far
unmentioned---renormalization prescription, such that eq.~\eqref{eq:hadMatch}
remains well-defined. For the employed Wilson fermion action, the quark mass
$\mbar_j$ can be obtained, on the lattice up to a finite renormalization, from
the non-singlet axial Ward identity, or current quark mass
\def\AWI{^{\rm\tiny awi}}
\begin{align}\label{eq:AWImass}
   \mbar_{j}\AWI(\mu) 
            &=  \lim_{a/L\to 0} \left[ \frac{\ZA(g_0^2)}{\ZP(g_0^2,a\mu)} m_{j}\AWI    \right]_{\mu=L^{-1}}  \;, &
   \partial_\mu \left[ \psibar_{j} \gamma_\mu\gamma_5  \psi_{j'}(x) \right] 
            &= (m_{j}\AWI+m_{j'}\AWI) \, \psibar_{j}(x) \gamma_5  \psi_{j'}(x)        \;.
\end{align}
The bare mass $m_j\AWI\equiv m_j\AWI(g_0^2,\{am_{0,i}\})$ still depends on the
bare parameters of the action, cf. eq.~\eqref{eq:QCDact}. From the previous
discussions the connection to the mass-SSF and notation should become evident
\begin{align}\label{eq:ratioAWImass}
   \sigP(s,u) \equiv  \frac{\mbar_{j}\AWI(L)}{\mbar_{j}\AWI(sL)} 
            &=  \lim_{a/L\to 0} \left[ \frac{\ZP(g_0^2,sL/a)}{\ZP(g_0^2,L/a)} \right]_{\gbsq(L)=u,m=0}  \;, 
\end{align}
cf. eqs.~\eqref{eq:ssf} and \eqref{eq:SSF}. $\ZP$ is the multiplicative
renormalization factor of the non-singlet pseudoscalar current, which itself is
determined from a SF renormalization condition as in~\cite{Capitani:1998mq}.

In Figure~\ref{fig:ssf} (left panel) we show preliminary results for some
continuum extrapolations as in eq.~\eqref{eq:ratioAWImass} with $s=2$. The
corresponding continuum SSF data points are then plotted against the initially
fixed target couplings $u$ for both the SF- and GF-coupling schemes (right
panel). Note that these values of $u$ can be chosen at will but should cover
the range of interest well enough in order to reach the desired precision.
Typical fit ansaetze as shown in Fig.~\ref{fig:ssf} are polynomials like 
$\sigP(u)=1+p_{\rm LO} u + p_2 u^2 + \cdots$ or Pad\'e approximants with
the correct, leading (universal) asymptotics fixed.  With such a given smooth
interpolation formula one is then able to recursively construct the running
factor and its uncertainty in steps of $s=2$ as summarised
in~\eqref{eq:running}, i.e., the two ratios in~\eqref{eq:masterF} become
\begin{align}\label{eq:SSFrunning}
     \frac{\mbar(\mu_a)}{\mbar(\mu_b)} &= \prod_{k=1}^{N_s} \sigP(u_k)  \;, & &
       \begin{cases} 
           u_{k+1}     = \sigma(u_{k}) \\
           u_1 \quad\! = \gbsq(\mu_a)  \\
       \end{cases} \!\!\!\!, &
       s_{ab} &\equiv \frac{\mu_a}{\mu_b} = \frac{L_b}{L_a} = 2^{N_s} \;.
\end{align}
Accordingly, we can identify the switching scale $\lswi$ appearing in
eq.~\eqref{eq:masterF} with the scale $2L_0$ of~\eqref{eq:u_match}, which we
inherit together with the coupling-SSF $\sigma(u)$ from the complementary
determination of the running
coupling~\cite{Brida:2016flw,DallaBrida:2016kgh,Bruno:2016gvs,DallaBrida:12Conf2016,Sint:12Conf2016}. This concludes
the computation of~\eqref{eq:masterF} and we return to the determination of
eq.~\eqref{eq:hadMatch}, which we can now rewrite as 
\begin{align}\label{eq:ZM}
        M_j  &= \frac{M}{\mbar(\muhad)} \times \lim_{a/L\to 0} \left[\frac{\ZA(g_0^2)}{\ZP(g_0^2,a\muhad)} m_{j}\AWI \right] 
              \equiv \lim_{a/L\to 0} \left[ \ZM(g_0^2) \,m_{j}\AWI(g_0^2,\{am_{0,i}\}) \right] \;.
\end{align}
The disappearance of the hadronic scale $\muhad$ reflects the
scale-independence of $M_j$, which itself is connected to the current quark
mass $m_j\AWI$ only by a scale-independent renormalization $\ZM$. Hence, by determining
$\ZM$ the renormalization problem is fully solved non-perturbatively using the
massless (intermediate) SF scheme.  While the costs of simulating the
Schr\"odinger functional setup are good to control in general, towards larger
physical volumes the simulations get more and more involved and thus set a
natural limit on the values the hadronic scale can take. At the moment we are
exploring different ways to fix the value of $a\muhad$ in order to determine
the $\ZP(g_0^2,a\muhad)$, and thus $\ZM$, with high precision. \newline
As the reader may have noticed, forcing $s_{ab}$ in eq.~\eqref{eq:SSFrunning}
to be a multiple of $2$ is a quite stringent condition, especially as we
already had to fix the scheme-switching scale in~\eqref{eq:masterF}. Thanks to
the high statistical accuracy and the achieved control of a few systematic
effects which are propagated into the final error, we are able to lift this
restriction for the first time. By reassessing eqs.~\eqref{eq:ssf}
and~\eqref{eq:SSF}, one notices that with the data points at hand and a given
parameterisation of the $\beta$-function, as shown in Figure~\ref{fig:betas}
and taken from~\cite{Brida:2016flw,DallaBrida:2016kgh}, only a well motivated
ansatz for $\tau(g)$ is required for a sensible least square minimisation. We
present a preliminary but already very encouraging result in
Figure~\ref{fig:tau}, again for both the SF- and GF-running coupling schemes.
It should be noted that this result is actually independent of the originally
chosen scale-change factor $s=2$. As a non-trivial cross-check we can use the 
resultant to numerically reconstruct $\sigP(u)$ for arbitrary $u$ from
eq.~\eqref{eq:ssf}. This is shown in Figure~\ref{fig:ssf} and coincides very
well with the original estimate. At the time of this conference we were still
accumulating statistics at the two strongest renormalized couplings in use.
Now we are finalising our analysis by also employing global fit ansaetze in
various ways and include some yet missing correlations between our data, which
is going to be published soon. Accordingly, we here refrain from quoting any
quantitative numbers.

Compared to the aforementioned simulations, the $m_j\AWI$ estimates
in~\eqref{eq:ZM} are obtained from large-scale lattice simulations that define
the accessible low-energy window in the first place. We will employ $\Nf=2+1$
ensembles jointly produced by the Coordinated Lattice Simulations (CLS)
effort~\cite{Bruno:2014jqa}, with which we share the lattice discretization in
use. Although there have been many advances in lattice QCD, the accessible
parameters $am_{0,i}$, in general, do not correspond directly to the set of
parameters at which $m_\pi$ and $m_{\rm K}$ take their physical values for
different values of the cutoffs (even after correcting for isospin splitting
and QED effects).  Beside setting the physical scale~\cite{Bruno:2016plf}, one
thus still has to apply chiral perturbation theory. In that respect,
eq.~\eqref{eq:ZM} is a minor simplification in our discussion, but nevertheless
a step that has to be carefully evaluated in the near future.

\section{Conclusions}

We have presented a full strategy to determine renormalized quark masses from
first principle lattice calculations in three-flavour QCD. Compared to other
approaches in the community, we separate and thus disentangle the problem of
renormalization from large-scale lattice QCD simulations. In this way the
massless renormalization group running for the coupling and quark masses could
be mapped out non-perturbatively to high accuracy in the employed Schr\"odinger
functional scheme, and a connection to other schemes can be easily established
via renormalization group invariant quantities $\{\Lambda,M_i\}_{\Nf}$. This
strategy fully circumvents the unanswerable question regarding the
applicability of perturbation theory at parametrically large values of the
strong coupling. Furthermore, it provides a quantifiable and systematically
improvable uncertainty at each stage of the calculation, something
perturbation theory cannot really provide.

For the first time two different non-perturbative schemes, based on the
Schr\"odinger functional coupling and the younger Gradient flow coupling, have
been used and matched non-perturbatively. This specific combination has been
chosen for the avail of improving on statistical and systematic uncertainties.
Also the non-perturbative determination of the continuum quark mass anomalous 
dimension is a unique achievement for QCD.

The prospects for determining fundamental parameters of QCD valuable for high
precision physics are excellent. To obtain an even more realistic approximation
of nature this study can be extended to cross the charm flavour threshold
$(\Nf=3\to\Nf=4)$ non-perturbatively at some point in the future. We finally
remark that also the renormalization and scale dependence of the tensor
current---relevant in rare meson decays (especially B decays), and studies of the
neutron electric dipole moment---is being computed along the same
lines~\cite{Fritzsch:2015lvo}.

\section*{Acknowledgments}
\small

The simulations were jointly performed on the Altamira HPC facility, on
Finisterrae-2 at CESGA, and on the GALILEO supercomputer at CINECA (INFN
agreement). We thankfully acknowledge the computer resources and technical
support provided by the University of Cantabria (IFCA), CESGA and CINECA.

P.F. acknowledges financial support from the Spanish MINECO's ``Centro de
Excelencia Severo Ochoa'' Programme under grant SEV-2012-0249, as well
as from the grant FPA2015-68541-P (MINECO/FEDER).

\bibliography{mainbib}

\begin{thebibliography}{32}

\bibitem{Brida:2016flw}
M.~Dalla~Brida, P.~Fritzsch, T.~Korzec, A.~Ramos, S.~Sint, R.~Sommer (ALPHA),
  Phys. Rev. Lett. \textbf{117}, 182001 (2016), \texttt{1604.06193}

\bibitem{PDG:2014}
K.~Olive et~al. (Particle Data Group), Chinese Physics C \textbf{38}, 090001
  (2014)

\bibitem{DallaBrida:2016kgh}
M.~Dalla~Brida, P.~Fritzsch, T.~Korzec, A.~Ramos, S.~Sint, R.~Sommer (ALPHA)
  (2016), \texttt{1607.06423}

\bibitem{Sint:12Conf2016}
S.~Sint (ALPHA), EPJ Web of Conferences \textbf{CONF12} (2016), this conference

\bibitem{DallaBrida:12Conf2016}
M.~Dalla~Brida (ALPHA), EPJ Web of Conferences \textbf{CONF12} (2016), this
  conference

\bibitem{Gross:1973id}
D.J. Gross, F.~Wilczek, Phys.Rev.Lett. \textbf{30}, 1343 (1973)

\bibitem{Politzer:1973fx}
H.D. Politzer, Phys.Rev.Lett. \textbf{30}, 1346 (1973)

\bibitem{Caswell:1974gg}
W.E. Caswell, Phys.Rev.Lett. \textbf{33}, 244 (1974)

\bibitem{Tarasov:1980au}
O.~Tarasov, A.~Vladimirov, A.Y. Zharkov, Phys.Lett. \textbf{B93}, 429 (1980)

\bibitem{vanRitbergen:1997va}
T.~van Ritbergen, J.~Vermaseren, S.~Larin, Phys.Lett. \textbf{B400}, 379
  (1997), \texttt{hep-ph/9701390}

\bibitem{Czakon:2004bu}
M.~Czakon, Nucl.Phys. \textbf{B710}, 485 (2005), \texttt{hep-ph/0411261}

\bibitem{Baikov:2016tgj}
P.A. Baikov, K.G. Chetyrkin, J.H. Kühn (2016), \texttt{1606.08659}

\bibitem{Tarrach:1980up}
R.~Tarrach, Nucl.Phys. \textbf{B183}, 384 (1981)

\bibitem{Tarasov:1982gk}
O.V. Tarasov (1982), in Russian, \texttt{preprint JINR P2-82-900}

\bibitem{Chetyrkin:1997dh}
K.~Chetyrkin, Phys.Lett. \textbf{B404}, 161 (1997), \texttt{hep-ph/9703278}

\bibitem{Vermaseren:1997fq}
J.~Vermaseren, S.~Larin, T.~van Ritbergen, Phys.Lett. \textbf{B405}, 327
  (1997), \texttt{hep-ph/9703284}

\bibitem{Baikov:2014qja}
P.A. Baikov, K.G. Chetyrkin, J.H. Kühn, JHEP \textbf{10}, 076 (2014),
  \texttt{1402.6611}

\bibitem{Luscher:1993gh}
M.~L{\"u}scher, R.~Sommer, P.~Weisz, U.~Wolff, Nucl.Phys. \textbf{B413}, 481
  (1994), \texttt{hep-lat/9309005}

\bibitem{Capitani:1998mq}
S.~Capitani, M.~L{\"u}scher, R.~Sommer, H.~Wittig (ALPHA), Nucl.Phys.
  \textbf{B544}, 669 (1999), \texttt{hep-lat/9810063}

\bibitem{Luscher:1992an}
M.~L{\"u}scher, R.~Narayanan, P.~Weisz, U.~Wolff, Nucl.Phys. \textbf{B384}, 168
  (1992), \texttt{hep-lat/9207009}

\bibitem{Sint:1993un}
S.~Sint, Nucl.Phys. \textbf{B421}, 135 (1994), \texttt{hep-lat/9312079}

\bibitem{Sint:1995rb}
S.~Sint, Nucl.Phys. \textbf{B451}, 416 (1995), \texttt{hep-lat/9504005}

\bibitem{Sommer:2015kza}
R.~Sommer, U.~Wolff (ALPHA), Nucl.Part.Phys.Proc. \textbf{261-262}, 155 (2015),
  \texttt{1501.01861}

\bibitem{Bruno:2016gvs}
M.~Bruno et~al. (ALPHA), \emph{{The determination of $\alpha_s$ by the ALPHA
  collaboration}}, in \emph{{Proceedings, 6th Workshop on Theory, Phenomenology
  and Experiments in Flavour Physics : Interplay of Flavour Physics with
  electroweak symmetry breaking. (Capri 2016): Anacapri, Capri, Italy, June
  11-13, 2016}} (2017), Vol. 285-286, pp. 132--138, \texttt{1611.05750}

\bibitem{Fritzsch:2013je}
P.~Fritzsch, A.~Ramos, JHEP \textbf{1310}, 008 (2013), \texttt{1301.4388}

\bibitem{Ramos:2015baa}
A.~Ramos, S.~Sint, Eur. Phys. J. \textbf{C76}, 15 (2016), \texttt{1508.05552}

\bibitem{Sint:1998iq}
S.~Sint, P.~Weisz (ALPHA), Nucl.Phys. \textbf{B545}, 529 (1999),
  \texttt{hep-lat/9808013}

\bibitem{Campos:2015fka}
I.~Campos et~al., PoS \textbf{LATTICE2015}, 249 (2016), \texttt{1508.06939}

\bibitem{Lattice16}
I.~Campos et~al., PoS \textbf{LATTICE2016}, 210 (2016)

\bibitem{Bruno:2014jqa}
M.~Bruno et~al., JHEP \textbf{1502}, 043 (2015), \texttt{1411.3982}

\bibitem{Bruno:2016plf}
M.~Bruno, T.~Korzec, S.~Schaefer (2016), \texttt{1608.08900}

\bibitem{Fritzsch:2015lvo}
P.~Fritzsch, C.~Pena, D.~Preti, PoS \textbf{LATTICE2015}, 250 (2016),
  \texttt{1511.05024}

\end{thebibliography}

\end{document}